\documentclass[twocolumn]{revtex4-2}
\usepackage{graphicx,epstopdf}
\usepackage{graphicx,booktabs,array}
\usepackage{epsfig}
\usepackage{amsmath}
\usepackage{amsfonts}
\usepackage{float}
\usepackage{amssymb}
\usepackage{amssymb}
\usepackage[colorlinks=true
,urlcolor=blue
,citecolor=blue
,linkcolor=blue
,pagecolor=blue
,linktocpage=true
,pdfproducer=medialab
]{hyperref}
\usepackage{subfigure}
\usepackage{longtable}
\usepackage{multirow}
\usepackage{notoccite}
\usepackage[nodisplayskipstretch]{setspace}
\usepackage{placeins}
\graphicspath{{./Diagrams/}}
\def\be{\begin{equation}}
\def\ee{\end{equation}}
\def\bea{\begin{eqnarray}}
\def\eea{\end{eqnarray}}

\def\lsim{\raise0.3ex\hbox{$\;<$\kern-0.75em\raise-1.1ex\hbox{$\sim\;$}}}
\def\gsim{\raise0.3ex\hbox{$\;>$\kern-0.75em\raise-1.1ex\hbox{$\sim\;$}}}

\usepackage{pstricks}
\usepackage{epsfig}
\usepackage{xcolor}
\begin{document}
\title{Resolving $R_{D}$ and $R_{D^*}$ Anomalies in Adjoint SU(5)}
\author{
A. Ismael$^{1,2}$ and S. Khalil$^2$}
\affiliation{
$^1$Physics Department, Faculty of Science, Ain Shams University, Cairo 11566, Egypt.}
\affiliation{$^2$Center for Fundamental Physics, Zewail City of Science and
Technology, 6th of October City, Giza 12578, Egypt.}
\date{\today}

\begin{abstract}
We investigate the $R_{D}$ and $R_{D^*}$   anomalies in the context of non-minimal $SU(5)$, where Higgs sector is extended by
adjoint 45-dimensional multiplet. One of the light spectrum of this model could be the scalar triplet leptoquark that is contained in this multiplet.
We demonstrate that this particular scalar leptogquark mediation of the transition $b \to c \tau \nu$ is capable of simultaneously accounting for both $R_{D}$ and $R_{D^*}$ anomalies.	
We further emphasize that another Yukawa coupling controls its contribution to $b \to s \ell^+ \ell^-$, ensuring that $R_K$ and $R_{K^*}$ remain consistent with the standard model predictions.	

\end{abstract}
\maketitle

\section{Introduction}
Semileptonic decays $B \to \{D,D^*\} \tau \nu$  have received a lot of attention in recent years because they provide a good opportunity to test the Standard Model (SM) and look for possible new physics beyond. Recent intriguing measurements of $R_{D,D^*}$ by  BaBar \cite{PhysRevLett.109.101802, PhysRevD.88.072012}, Belle \cite{PhysRevD.92.072014, PhysRevLett.118.211801, PhysRevD.97.012004, PhysRevLett.124.161803}, and LHCb collaborations \cite{https://indico.cern.ch/event/1187939/} are significant hints of new physics that violate lepton flavor universality. The ratios $R_{D,D^*}$ are  defined by
\begin{align}
R_{D^*,D} \equiv \frac{BR(B_q\rightarrow \{D^*,D\}
\tau\nu)}{BR(B_q\rightarrow \{D^*,D\} l\nu )}\,,
\end{align}
where $l=e,\mu$. The current experimental averages of $R_D$ and
$R_{D^*}$ are given by \cite{HFLAV}
\begin{align}
R_D &= 0.339 \pm  0.026 \pm  0.014\,, \label{eq:current-data-2}\\
R_{D^*}   &= 0.295 \pm 0.010 \pm 0.010\,.
\label{eq:current-data-1}
\end{align}
However, the SM predictions are given as follows: \cite{Bigi:2016mdz,Gambino:2019sif,Bordone:2019vic} 
\begin{align}
R_D^{\mathrm{SM}}   &= 0.298 \pm 0.004\,, \\
R_{D^*}^{\mathrm{SM}} &= 0.254 \pm 0.005\,.
\end{align}
This shows that the measured $R_D$ and $R_{D^*}$ results deviate from the SM expectations by $1.9\sigma$ and $3.2\sigma$, respectively. On the other hand, the LHCb recently announced new results for the ratios 
\bea 
R_{K}=\frac{\text{BR}(B^+\to K^+ \mu^+\mu^-)}{\text{BR}(B^+\to K^+ e^+ e^-)},\\
R_{K^*}=\frac{\text{BR}(B^0\to K^{*0} \mu^+\mu^-)}{\text{BR}(B^0\to K^{*0} e^+ e^-)}.
\eea
It has been reported that $R_K$ and $R_{K*}$ are given for two dilepton invariant mass-squared bins by
~\cite{https://doi.org/10.48550/arxiv.2212.09152,https://doi.org/10.48550/arxiv.2212.09153}
\begin{align}
\label{Eq:RKRKsexp1}
{\rm Low}-q^2~~~~ 
&\left\{ \begin{array}{l}
R_{K} = 0.994~ ^{+0.09}_{-0.082} ~({\rm stat}) ~ ^{+0.027}_{-0.029} ~({\rm syst})\nonumber\\ \\
R_{K^*} = 0.927~ ^{+0.0933}_{-0.087} ~({\rm stat}) ~ ^{+0.034}_{-0.033} ~({\rm syst})
 \end{array} \right.\nonumber\\ \\
{\rm Central}-q^2~~~~
&\left\{ \begin{array}{l} 
R_{K} = 0.949~ ^{+0.042}_{-0.041} ~({\rm stat}) ~ ^{+0.023}_{-0.023} ~({\rm syst})\nonumber\\ \\
R_{K^*} = 1.027~ ^{+0.072}_{-0.068} ~({\rm stat}) ~ ^{+0.027}_{-0.027} ~({\rm syst})\nonumber
 \end{array} \right.\nonumber
\label{Eq:RKRKsexp2}
\end{align}
These measurements are consistent with the SM predictions: $R_{K,K^*} \simeq 1$~\cite{RKSM}. As a result, they would impose sever constraints on any new physics contributions that could lead to lepton flavor non-universality.

In this paper, we argue that the scalar triplet leptoquark within the adjoint SU(5) framework can account for the discrepancy between  $R_{D,D^*}$ experimental results and SM expectations, while preserving $R_{K,K^{*}}^{\rm SM}$ results. The Adjoint $SU(5)$ is the simplest extension of minimal $SU(5)$ Grand Unified Theory (GUT), in which the Higgs sector is extended by a 45-dimensional multiplet ($45_H$). As is well known, minimal $SU(5)$ has a number of serious problems, such as the incorrect prediction for the fermion mass relation: $m_{\mu(e)} = m_{s(d)}$. One possible solution to some of these flaws is to introduce an extra $45_H$. The scalar triplet is one of the $45_H$ components, with the following $(3^*,2,-7/6)$ representation under the SM gauge group. Because of its special interactions with quarks and leptons, this scalar triplet, which is a leptoquark type particle, does not contribute to proton decay, as explained in \cite{LPTOdec}. This distinguishes $SU(5)$ scalar triplet from previous leptoquark scenarios discussed in the literature. \cite{LQ1,LQ2,LQ3,LQ4}. Although the scalar letptoquark contributes to the semileptonic decays $b \to c \tau \nu$ at the tree level, it is still subdominant because the leptoquark's mass is quite heavy of order TeV, which is sufficient to account for the given $\sim 10\%$ discrepancy. Controlling the contribution of scalar leptoquarks to the $b \to s \ell^+ \ell^-$ can be accomplished by constraining one of the free Yukawa couplings.

The paper is organized as follows. In section 2 we introduce the $SU(5)$ scalar leptoquark and its associated interactions, emphasizing that it does not contribute to proton decay but can play important role in the following decays: $b \to c \tau \nu$ and $b \to s \ell^+ \ell^-$. 
Section 3 is devoted to anlayzing the new contribution of our scalar leptoquark to $R_{D,D^*}$. 
$R_{K,K^*}$ analysis is discussed in section 4. Finally our conclusions and prospects are give in section 5.

%
\section{Scalar Leptoquark in Adjoint $SU(5)$}

As previously advocated, extending the Higgs sector of $SU(5)$  by $45_H$ helps to solve some of the problems that this simple example of GUT model faces \cite{GUT, gut, khalil, KHALIL}. The $45_H$ transforms under the SM gauge as%
\bea %
45_H &= &(8,2)_{1/2}\oplus (1,2)_{1/2}\oplus (3,1)_{-1/3} \oplus (3,3)_{-1/3} \nonumber\\
& \oplus&  (6^*,1){-1/3}\oplus (3^*,2)_{-7/6}\oplus
(3^*,1)_{4/3}.%
\eea%
It also satisfies the following constraints: $45^{\alpha
\beta}_\gamma = - 45^{\beta \alpha}_\gamma$ and $\sum_\alpha^5
(45)^{\alpha \beta}_\alpha =0$. Through non-vanishing Vacuum Expectation Values (VEVs) of $5_H$
and $45_H$: $\langle 5_H \rangle = v_5,\langle 45_H \rangle^{15}_1 = \langle 45_H \rangle^{25}_2 =
\langle 45_H \rangle^{35}_3 = v_{45}, \langle 45_H \rangle^{45}_4 = -3 v_{45}$, the electroweak symmetry $SU(2)_L \times U(1)_Y$ is spontaneously broken into
$U(1)_{em}$.

The $45_H$ scalar triplets are defined as:
\bea 
&&  (3^*,2)^{ij}_{c\ -7/6} \equiv (45_H)^{ij}_c \equiv \Phi^{ij}_c ,
\\\nonumber &&  (3^*,1)^{a b}_{k\ 4/3} \equiv (45_H)^{ab}_k \equiv  \Phi^{ab}_k ,
\\\nonumber && [(3,1)^{ib}_c \oplus (3,3)^{ib}_c]_{-1/3} \equiv (45_H)^{ib}_c \equiv \Phi^{ib}_c .
\eea %
It has been emphasized \cite{LPTOdec} that while the scalar triplets $\Phi^{ab}_k$ and $ \Phi^{ib}_c $ contribute to the proton decay and they must be superheavy,  the scalar triplet 
$\Phi^{ij}_c$ does not. It has no interaction terms that would cause proton decay.  By writing $\Phi^{ij}_c$ as $(\phi^i_1, \phi^i_2)^T$, one can demonstrate that the scalar triplet has the following peculiar interactions:
\bea
{\cal L}\!&\!=\!&\! 2 Y^2_{AB}  e^{T}_A C u^c_{B i}  \phi^{i1*}  \!+\! 4 (Y^4_{AB}\!-\! Y^4_{BA} ) u^{iT}_A C e^{c}_B \phi_{i1} \nonumber\\
\!&\!-\!&\! 2 Y^2_{AB} \nu^{T}_A C u^c_{B i}  \phi^{i2*} \!+\! 4 (Y^4_{AB}\!-\! Y^4_{BA} ) d^{iT}_A C e^{c}_B \phi_{i2}.~~~
\eea
The first two interaction terms would imply the decay of  $b \to s \ell^+ \ell^-$ through scalar triplet leptoquark $\phi^{i1}$ mediation, while the last two interaction terms  
clearly account for the decay $b \to c \tau \nu$ via scalar triplet leptoquark $\phi^{i2}$ mediation. These terms can be written as 
\be
\!\!\!{\cal L} = 2Y^2_{AB}\bar{u}_{Bi}P_{L}\nu_{A}\phi^{i2*}-4Y_{AB}^{4'}\bar{e}_{B}P_Ld^{i}_{A}\phi_{i2}+h.c.,
\ee 
where we used $C^T=-C$ and $\bar{\Psi}=\Psi_{L}^{cT}$, and define $Y^{4'}_{AB} \equiv(Y^4_{AB}-Y^4_{BA})$.  In the mass eignestate basis, where 
$$d_A \to V^{CKM}_{AB} d_B,~~  \nu_A \to V^{\rm PMNS}_{AB} \nu_B,~~ u_A \to u_A,~~   e_A \to e_A,$$ 
the above Lagrangian takes the form:
\bea
\mathcal{L} &=& 2Y^2_{ AB}\bar{u'}_{Bi}P_{L}V_{ AK}^{\rm PMNS}\nu'_{k}\phi^{i2*} - 4Y^{4'}_{ AB}\bar{e'}_{B}P_{L}V^{\rm CKM}_{ AK}d'_{K}\phi_{i2}\!\nonumber\\
&+&\!h.c.
\eea

In this regards, the amplitude of $b \to c \tau \nu$ transition is given by 
\begin{eqnarray}
\mathcal{M}\!&\!=\!&\!-\frac{8Y^{4'}_{13}V_{13}^{\rm CKM}}{M^2_{\phi}}\Big[\frac{1}{2}(\bar{u}_\tau P_L v_{\nu_{\tau}})(\bar{u}_C P_L u_b)\nonumber\\
&+&\frac{1}{8}(\bar{u}_\tau \sigma^{\mu\nu}P_L v_{\nu_{\tau}})(\bar{u}_C P_L\sigma^{\mu\nu} u_b)  \times\Big(Y^2_{12}V_{13}^{\rm PMNS}\\
&\!+\!&\!Y_{22}^{2}V_{23}^{\rm PMNS}\!+\!Y^2_{32}V_{33}^{\rm PMNS}\Big) \Big]\!+\!\Big(Y^{4'}_{13}V_{13}^{\rm CKM}\!\rightarrow \!Y_{23}^{4'}V_{23}^{\rm CKM}\Big).\nonumber
\label{amplitude}
\end{eqnarray}
Because $V^{CKM}_{13}$ and $V^{CKM}_{23}$ are so small ($10^{-3}$ and $10^{-2}$, respectively), the amplitude of $b \to c \tau \nu$ is essentially determined by the leptoquark masses $M_{\phi}, Y^2_{22}, Y^2_{32}$, and $Y^{4'}_{13}$.

%
\section{$SU(5)$ Leptoquark contribution to $R_{D,D^*}$}

The general expression of the effective Hamiltonian for $b\rightarrow cl\bar{\nu_{l}}$ can be written as \cite{Eff}
\begin{eqnarray}
{\mathcal{H}}_{\mathrm{eff}} &=& \frac{4G_{F}V_{cb}}{\sqrt{2}}\Big{[}
(1+g_{V_L})[\bar{c}\gamma_{\mu}P_{L}b][\bar{l}\gamma_{\mu}P_{L}\nu_{l}]\nonumber\\
&+& g_{V_R} [\bar{c}\gamma_{\mu}P_{R}b][\bar{l}\gamma_{\mu}P_{L}
\nu_{l}]+ g_{S_L} [\bar{c}P_{L}b][\bar{l}P_{L}\nu_{l}]  \nonumber\\
 &+& g_{S_R} [\bar{c}P_{R}b][\bar{l}P_{L}\nu_{l}]+g_{T} [\bar{c}\sigma^{\mu\nu_\tau
}P_{L}b][\bar{l}\sigma_{\mu\nu}P_{L}\nu_{l}]\Big{]}\!,~~~
\end{eqnarray}
where $G_{F}$ is the Fermi coupling constant,  $V_{cb}$ is the Cabibbo-Kobayashi-Maskawa (CKM) matrix element between charm and bottom quarks while $P_{L/R}=(1\mp \gamma_{5})/2$.
Here, $g_i$ is defined as  $g_{i}=C_{i}^{\rm NP}/C^{\mathrm{SM}},~i\equiv V_L, V_R, S_L, S_R, T$, with $C^{\mathrm{SM}}=\frac{4G_{F}V_{cb}}{\sqrt{2}}$. 
Eq. \ref{amplitude} shows that $g_{V_L}=g_{V_R}=g_{S_R}=0$, whereas $g_{S_L}$ and $g_{T}$ are given by 
\be
g_{S_L}=-\frac{\sqrt{2} Z}{M^2_{\phi}G_F}, ~~~~~~  g_{S_T}=\frac{g_{SL}}{4} =-\frac{Z}{2\sqrt{2}M^2_{\phi}G_F},
\ee
with 
\bea
Z&=&\Big(Y^2_{12}V_{13}^{\rm PMNS}+Y_{22}^{2}V_{23}^{\rm PMNS}+Y^2_{32}V_{33}^{\rm PMNS}\Big)\nonumber\\
&&\Bigg(\frac{Y^{4'}_{13}V^{CKM}_{13}}{V_{23}^{CKM}}+Y^{4'}_{23}\Bigg)
\eea

Substituting with the SM parameters as well as the form factors involved in the definition of the matrix elements to their central values, one finds \cite{Numer}
\bea
R(D)&=&{R}(D)^{\mathrm{SM}}\Big[1+ 1.02|g_{SL}|^{2} +0.9 |g_{T}|^{2} \nonumber\\
&+&1.49 \operatorname{Re}[g_{SL}^{\ast}]+1.14\operatorname{Re}[g_{T}^{\ast}]\Big],\\
R({D^{\ast}}) &=&{R}(D^{\ast})^{\mathrm{SM}}\Big[1+ 0.04 |g_{SL}|^{2}+ 16.07|g_{T}|^{2}\nonumber\\
&-& 0.11\operatorname{Re}[g_{SL}^{\ast}]-5.12\operatorname{Re}[g_{T}^{\ast}]\Big].
\eea

A few remarks are in order. First, the $g_{S_L}$ and $g_T$ can be complex due to non-zero phases in $U^{\rm PMNS}$ as well as complex values of the Yukawa couplings $Y^2$ and $Y^{4'}$. Second, because the tree-level scalar leptoquark contributes to the branching ratio of the tauonic decay $B_c^- \to \tau^-
\bar{\nu}_{\tau}$, experimental constraints from this decay must be included in our analysis. 
The modified branching ratio ${\rm BR}(B_c^- \to \tau^- \bar{\nu}_{\tau})$ is given by \cite{Numer, BR, br}
\be 
\label{BRBc_TypeII}
\hspace{-0.2cm}{\rm BR}(B_c^- \!\to\! \tau^- \bar{\nu}_{\tau}) \!=\! {\rm BR}(B_c^- \!\to\!
\tau^- \bar{\nu}_{\tau})_{\text{SM}} \vert  1\!-\!4.065 g_{S_L}\vert^2\!, 
\ee
with ${\rm BR}(B^-_c \to \tau^-
\bar{\nu}_\tau)_{\rm SM} = (2.25 \pm 0.21)\times 10^{-2}$
\cite{Fleischer:2021yjo}. The experimental bound on
${\rm BR}(B^-_c \to \tau^- \bar{\nu}_\tau)$ varies from $\leq 10\%$ to $\leq 60\%$ \cite{Bardhan:2019ljo, Blanke:2018yud, Blanke:2019qrx, Fleischer:2021yjo}. Third, it is also worth noting that our type of scalar leptoquarks would not contribute to lepton flavor violation, like $\tau \to \mu \gamma$ or $B-\bar{B}$ mixing. Fourth, we impose the constraints of the $D^*$ and $\tau$ longitudinal polarizations that come from Belle experiment. Their expressions depend on the same Wilson coefficients affecting $R_D$ and $R_{D^{*}}$, which are written as \cite{Numer, br}
\bea
\frac{F_{L}^{D^{*}}}{F_{L,SM}^{D^{*}}}&=&\Big(\frac{R_{D^{*}}}{R_{D^{*}}^{SM}}\Big)^{-1}\Big[1+0.08|g_{SL}|^2+7.02|g_{T}|^2\nonumber\\
&-&0.24~Re[g_{SL}^{*}]-4.37~Re[g_{T}^{*}]\Big]\\
\frac{P_{\tau}^{D^{*}}}{P_{\tau,SM}^{D^{*}}}&=&\Big(\frac{R_{D^{*}}}{R_{D^{*}}^{SM}}\Big)^{-1}\Big[1-0.07|g_{SL}|^2-1.86|g_{T}|^2\nonumber\\
&+&0.22~Re[g_{SL}^{*}]-3.37~Re[g_{T}^{*}]\Big]
\eea
The experimental values of $F_{L}^{D}$ and $P_{\tau}^{D^{*}}$ are given by $0.60\pm 0.08\pm 0.035$ \cite{BelleI} and $-0.38\pm 0.51^{+0.21}_{-0.16}$ \cite{PhysRevLett.118.211801, PhysRevD.97.012004, BelleII} respectively, whereas their SM predictions are $0.46\pm0.04$ \cite{BelleIII} and $-0.497 \pm 0.013$ \cite{BelleIV} Finally, running the coefficients $g_{S_L}$ and $g_{T}$ from the scale $\mu = 1 TeV$ to the scale $m_b = 4.2 GeV$ implies that \cite{runningI, runningII}:
\be \left(%
\begin{array}{c}
    g_{S_L}\\
   g_{T}  \\
   \end{array}%
\right)=\left(%
\begin{array}{cc}
 1.71 & 0  \\
0 & 1  \\
 \end{array}%
\right)\left(%
\begin{array}{c}
g_{S_L} (\mu = 1 TeV)\\
g_{T} (\mu = 1 TeV)\\
\end{array}%
\right).\ee

\begin{figure*}[!t]
\includegraphics[width=8cm,height=6cm,angle=0]{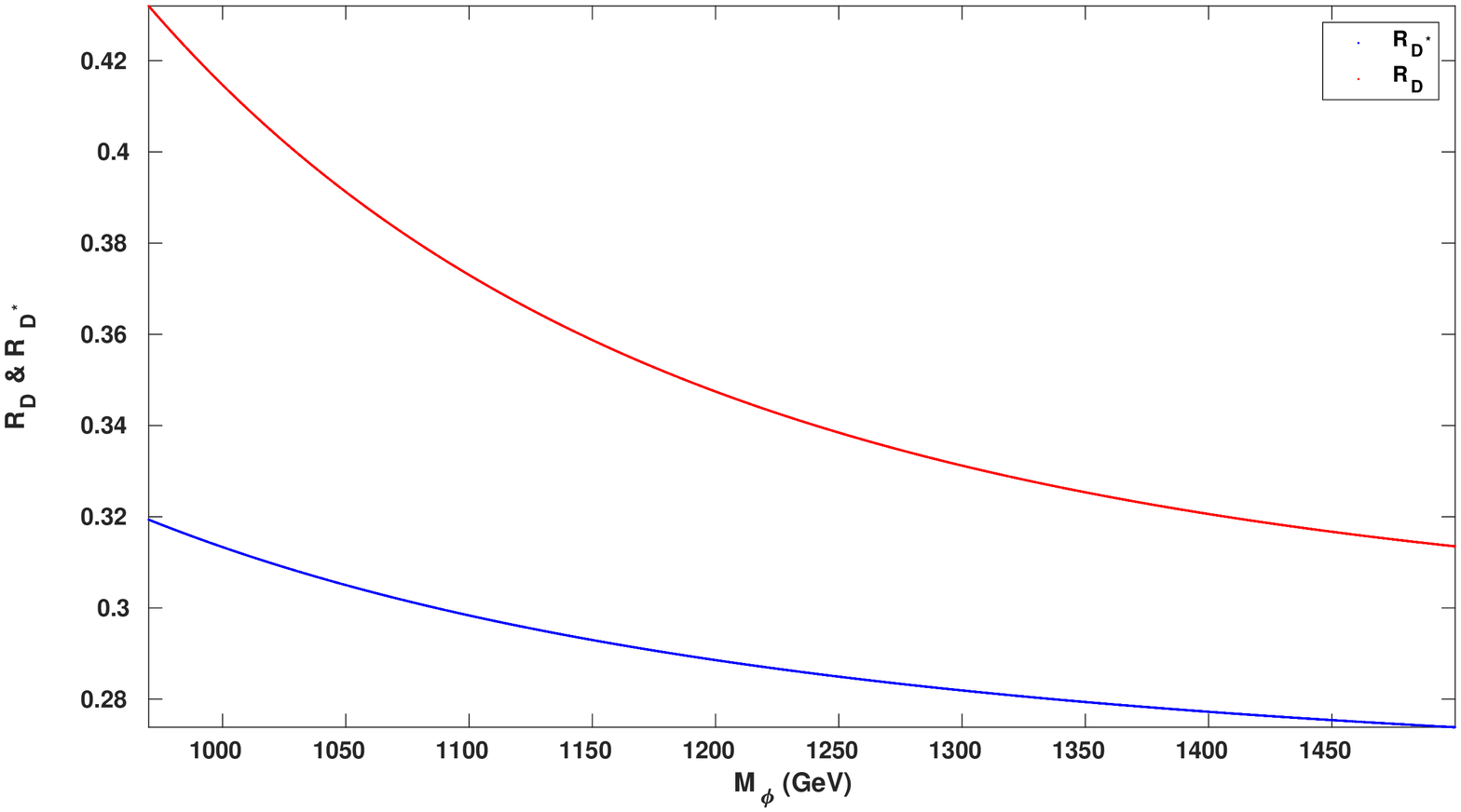}~~~~~~~~
\includegraphics[width=8cm,height=6cm,angle=0]{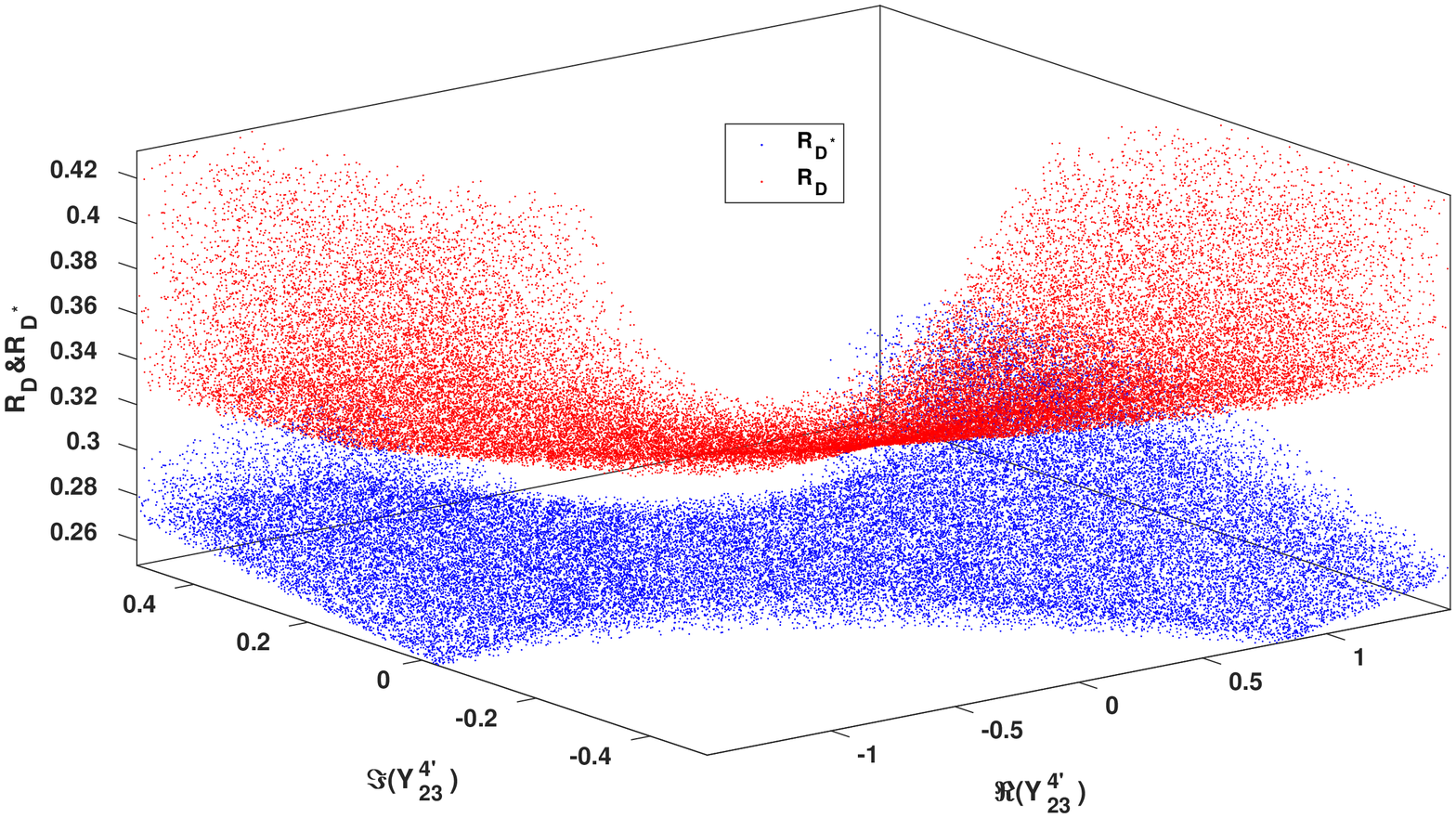}
\caption{$R_D$ and $R_{D^*}$ as function of the Letoquark mass and and real and imaginary parts of the Yukawa coupling $Y_{23}$. The other parameters are fixed as mentioned in the text. }
\label{fig1}
\end{figure*}

In the presence of the aforementioned experimental constraints, we performed a numerical analysis of $R_D$ and $R_{D*}$. In Fig. \ref{fig1}, we show the dependence of $R_D$ and $R_{D^*}$  on the most relevant parameters, which are the mass of leptoquark $M_{\phi}$ (left panel) and the real and imaginary parts of the Yukawa coupling $Y_{23}^{4'}$ (right panel).  The other parameters in these plots were set as follows: $Y^{2}_{12}=-1.5$, $Y^{2}_{22}=Y^{2}_{32}=Y^{4'}_{13}=1.5$. Furthermore, the coupling $Y_{23}^{4'}$ is fixed with $1.48+0.1 i$ in the plot of $R_D$ and $R_{D^*}$ versus $M_\phi$ (left panel), whereas in the 3D plot of $R_D$ and $R_{D*}$ versus real and imaginary parts of $Y_{23}^{4'}$ (right panel), the mass $M_\phi$ varies along the $[800,1500]$ GeV, while real and imaginary parts of $Y_{23}^{4'}$ vary along the $[-1.5,1.5]$ and $[-0.5,0.5]$, respectively.

\begin{figure*}[htb!]
\includegraphics[width=8cm,height=6cm,angle=0]{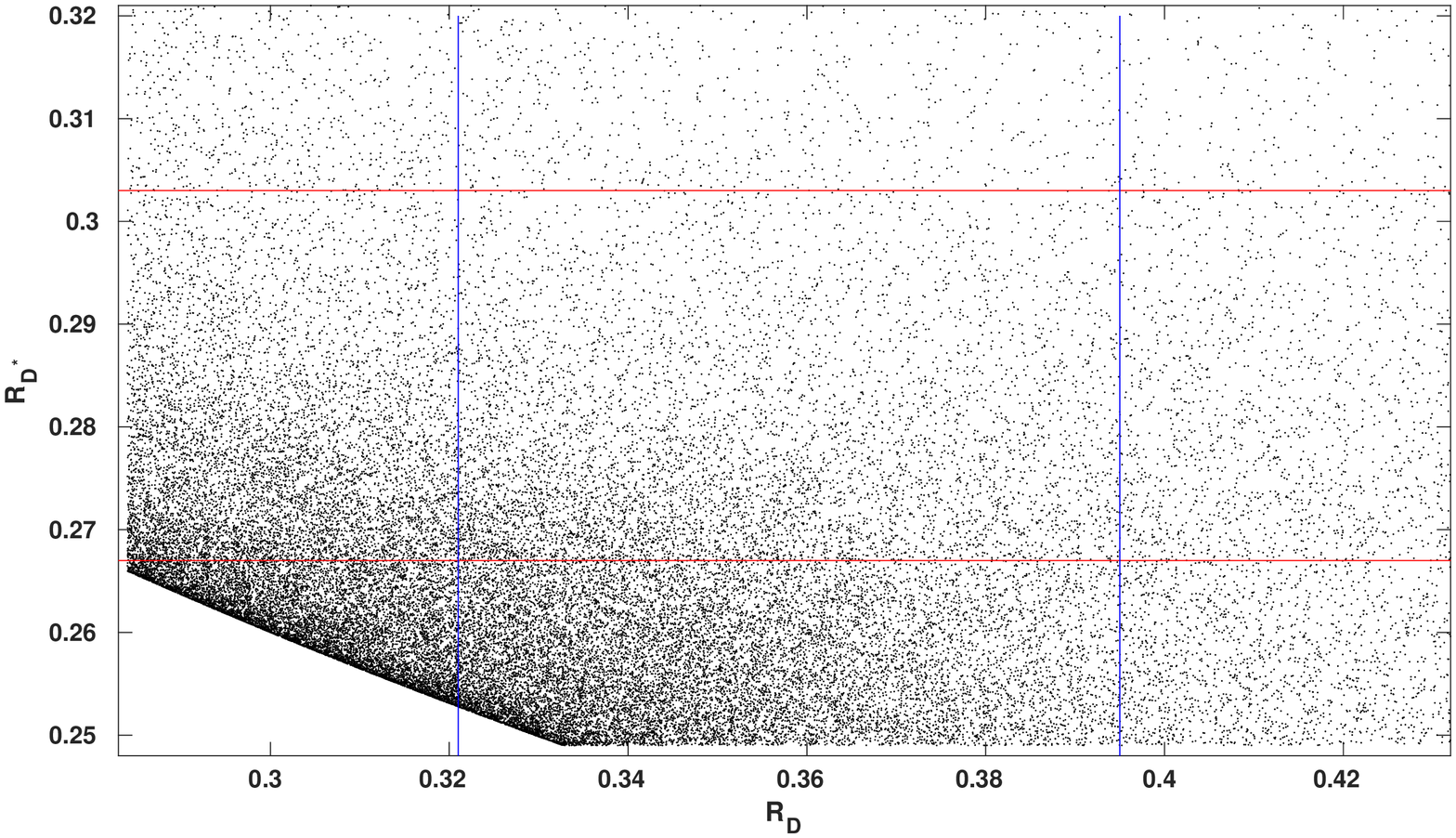}~~~~
\includegraphics[width=8cm,height=6cm,angle=0]{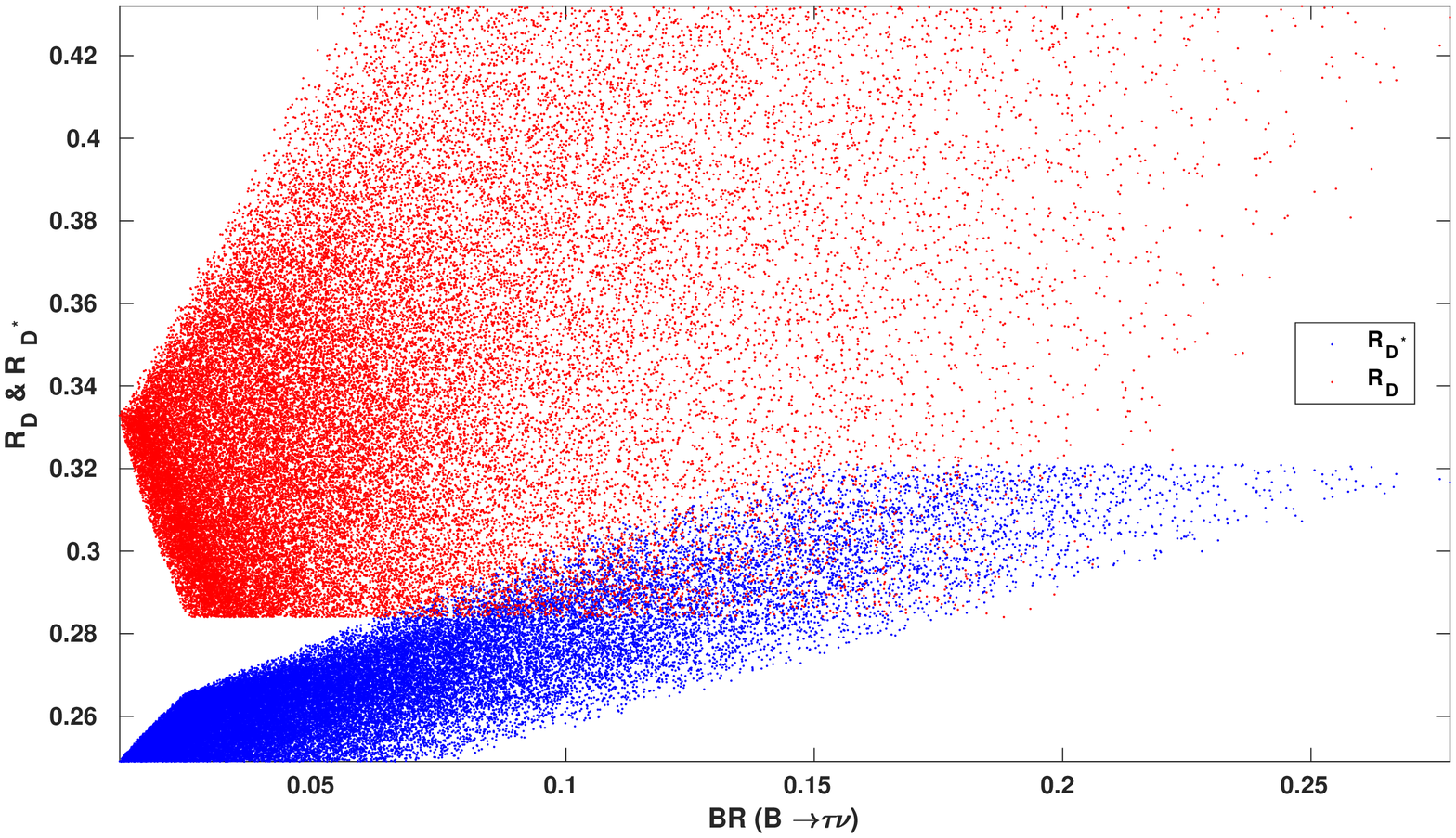}
\caption{The correlation between $R_D$ and $R_{D^*}$ (left) and between both $R_D$ and $R_{D^*}$ and ${\rm BR}(B^-_c \to \tau^- \bar{\nu}_\tau)$ (right). The scan is conducted over the regions of parameter space mentioned above. }
\label{fig2}
\end{figure*}

The correlation between $R_D$ and $R_{D^*}$  is shown in Fig. \ref{fig2}, left-panel, and the correlation between the constraints on the ${\rm BR}(B^-_c \to \tau^- \bar{\nu}_\tau)$ and $R_D$ and $R_{D*}$ is highlighted in the right-panel of this plot. The parameters are set in the same way as in the previous plots.

These plots show that in this class of models, both $R_D$ and $R_{D^*}$ can be significantly enhanced and lie within one sigma of the recent experimental limits, with scalar leptoquark masses of order one TeV, which is consistent with experimental constraints.
\section{$SU(5)$ Leptoquark contribution to $R_{K,K^*}$}
In this section, we show that, while the scalar leptoquark causes non-universality of lepton flavor in the process $B \to D \ell \nu$, it does not necessarily cause non-universality in the process $B \to K \ell^+ \ell^-$.
The Lagrangian that generates the $b \to s \ell^+\ell^-$ transition is given by
\begin{equation}
\mathcal{L}\!=\!-4Y^{4'}_{AB}\bar{e'}_B P_L V^{CKM}_{AK}d^{i'}_{K}\phi_{i2}\!-\!4Y^{4'}_{AB}\bar{d'^{i}}_K V^{CKM^*}_{AK}P_R~e^{'}_{B}\phi_{i2}^*.
\end{equation}
Thus, for $b\rightarrow s~\mu~\mu^+$, the Lagrangian is given as 
\bea
\mathcal{L}&\supset& -4Y^{4'}_{32}\bar{\mu '}P_L b^{i'}\phi_{i2}-4Y_{12}^{4'*}\bar{S^{i'}}V_{12}^{CKM^*}P_R~\mu^{'}\phi_{i2}^{*}\nonumber\\
&-&4Y_{32}^{4'*}\bar{S^{i'}}V_{32}^{CKM^*}P_R~\mu^{'}\phi_{i2}^{*},
\eea
 where $V_{13}^{CKM}\approx 0$ and $V_{33}^{CKM}\approx 1$ are assumed. Also, we may neglect $V_{32}^{CKM}$ respect $V_{12}^{CKM}$ (although we include all terms in our numerical calculations). Thus, the amplitude of this process is given by
 \begin{equation}
 \mathcal{M}=\frac{8 Y^{4'}_{32}Y^{4'*}_{12}V_{12}^{CKM^*}}{M^2_{\phi}}\big(\bar{U}_{s}\gamma_{\mu}P_LU_{b}\big)\big(\bar{U}_{\mu}\gamma^{\mu}P_LV_{\nu}\big).
\end{equation} 
We used the Fierz transformation identity
\begin{equation}
\big(\bar{U}_{s}P_RV_{\mu}\big)\big(\bar{U_{\mu}}P_LU_{b}\big)=\frac{1}{2}\big(\bar{U}_{s}\gamma_{\mu}P_LU_b\big)\big(\bar{U}_{\mu}\gamma^{\mu}P_Lv_{\mu}\big).
\end{equation}
As a result, the Wilson coefficient $C_{9}^{\mu}$ for  $b\rightarrow s~\mu~\mu^+$ process is written as
\begin{equation}
C_{9}^{\mu}(\Lambda)=\frac{8Y^{4'}_{32}Y^{4'*}_{12}V_{12}^{CKM^*}}{M^2_{\phi}}.
\end{equation}
where the scale $\Lambda\approx 1\mbox{TeV}$, and $C_{10}^{\mu}(\Lambda)=-C_{9}^{\mu}(\Lambda)$. On the other hand, the Lagrangian that generates the process $b\rightarrow s~e~e^+$ is given by
\begin{align}
\mathcal{L}=&-4Y_{31}^{4'}\bar{e^{'}}P_{L}b^{i'}\phi_{i2}-4Y_{21}^{4'*}\bar{S^{i'}} P_{R}e^{'}\phi_{i2}^{*}.
\end{align}
 After applying Fierz identity, the amplitude of $b\rightarrow s~e~e^+$ is given by
 \begin{equation}
 \hspace{-12mm}\mathcal{M}=8\frac{Y_{31}^{4'}Y_{21}^{4'*}}{M_{\phi}^2}\big(\bar{U}_{s}\gamma_{\mu}P_LU_{b}\big)\big(\bar{U}_{e}\gamma^{\mu}P_LV_{e}\big).
 \end{equation}
 Hence, the Wilson coefficient $C_{9}^{e}(\Lambda)$ for $b\rightarrow s~e~e^+$ will be
 \begin{equation}
 C_{9}^{e}(\Lambda)=8\frac{Y_{31}^{4'}Y_{21}^{4'*}}{M_{\phi}^2}.
 \end{equation}
Moreover, $C_{10}^{e}(\Lambda)=-C_{9}^{e}(\Lambda).$ The effective Hamiltonian $H_{eff}$ for $R_K$ process is given by
\begin{equation}
H_{eff}=\sum_{i}\big(C_{i}(\mu_{b})\mathcal{O}_i(\mu_b)+\tilde{C_i}(\mu_b)\tilde{\mathcal{O}}_i(\mu_b)\big).
\end{equation}
Through renormalization group equation (RGE), we obtain 
\begin{equation}
C_{9,10}^{e,\mu}(\Lambda)=1.2~C_{9,10}^{e,\mu}(\mu_b),
\end{equation}
where $\mathcal{O}_{i}(\mu_b)$ are $\Delta B=1$ transition operator, which is evaluated at the $m_b$ scale. $\tilde{C_i}(\mu_b),~\tilde{\mathcal{O}}_i(\mu_b)$ are obtained by replacing $L\leftrightarrow R$. The relevant operators that describe the $R_{k}$ and $R_{k^*}$ in our model are
\begin{equation}
\mathcal{O}_9=\big(\bar{s}\gamma_{\mu}P_L b\big)\big(\bar{l}\gamma^{\mu}l),~~\mathcal{O}_{10}=\big(\bar{s}\gamma_{\mu}P_L b\big)\big(\bar{l}\gamma^{\mu}\gamma_{5}l).
\end{equation}
The $R_{k}$ and $R_{k^*}$ expressions are written as
\begin{align}
R_k\approx&1+\Delta_{+},\\
R_{k^{*}}\approx&1+\Delta_{+}+p(\Delta_+-\Delta_-),
\end{align}
 where p is a function of $q^2_{min}$ and $q^2_{max}$ and $\Delta_{\pm}$ is given by
 \bea
 \Delta_{\pm}&=&\frac{2}{|C_9^{SM}|^2+|C_{10}^{SM}|^2}\Bigg[\Re\Big(C_{9}^{SM}(C_{9}^{NP,\mu}\pm\tilde{C_{9}}^{NP,\mu})^*\Big)\nonumber\\
 &+&\Re\Big(C_{10}^{SM}(C_{10}^{NP,\mu}\pm\tilde{C_{10}}^{NP,\mu})^*\Big)-(\mu\leftrightarrow e)\Bigg]
 \eea
 For our model, $\tilde{C}_{9,10}^{NP}=0$. Therefore, we obtain
 \begin{equation}
 \Delta_+=\Delta_-=\frac{2.4~\big(C_{9}^{SM}-C_{10}^{SM}\big)}{|C_{9}^{SM}|^2+|C_{10}^{SM}|^2}\Re\Big(C_9^{NP,\mu}(\mu_b)-C_9^{NP,e}(\mu_b)\Big)^*
 \end{equation}

It is worth mentioning that, whereas $R_{K,K^*}$ is essentially dependent on the couplings $Y^{4'}_{21}$ and $Y^{4'}_{32}$, $R_{D,D^*}$ is dependent on  $Y^2_{22}$, $Y^2_{33}$ and $Y^{4'}_{23}$. As a result, it is entirely possible to keep $R_{K,K^*}$ equal to the SM expectation, consistently with the new LHCb results, while leaving $R_{D,D^*}$ intact. To make $R_{K,K^*}$ close to one, $\Delta_+$ should be very small. This can be accomplished by having $Y^{4'}_{12} \ll 1$.

\section{Conclusions}
In this paper we have demonstrated that, in the presence of experimental constraints on flavor and lepton violation observables, measured values of  $R_D$ and $R_{D^*}$ within $1\sigma$ can be explained in non-minimal $SU(5)$ with adjoint $45$-dimensional Higgs multiplet. Enhancements for both $R_D$ and $R_{D^*}$ are made possible by a tree level transition of $b \to c \tau \nu$, which is mediated by the associated scalar leptoquark. We also emphasized that even though this leptoquark may contribute to $R_{K}$ and $R_{K^*}$, they remain independent of $R_D$ and $R_{D^*}$ enhancements because they are given in terms of different Yukawa couplings. As a result, their contributions can be easily suppressed, and $R_{K}$ and $R_{K^*}$ continue to be identical to SM predictions, which are consistent with the most recent LHCb data.

\section*{acknowledgements}
This work is partially supported by Science, Technology $\&$ Innovation Funding Authority (STDF) under grant number 37272.
\newpage

\end{document}